\documentclass[twocolumn,letterpaper,aps,prc,superscriptaddress,nofootinbib,showpacs,floatfix]{revtex4-1}
\usepackage{comment}
\usepackage{setspace}
\usepackage{color}
\usepackage{xcolor}
\usepackage{indentfirst}
\usepackage{xspace}
\usepackage{verbatim}
\usepackage{epstopdf}
\usepackage[export]{adjustbox}
\usepackage[T1]{fontenc}
\usepackage{float}
\usepackage{wrapfig}
\usepackage{graphicx}
\usepackage{subcaption}
\usepackage{diffcoeff}
\usepackage{amsmath}
\usepackage{amssymb}
\usepackage{amsthm}
\usepackage{array}
\usepackage{tabularx}
\usepackage[utf8]{inputenc}
\graphicspath{ {figures/} }
\usepackage{lineno}
{\large }

\usepackage{natbib}
\setcitestyle{square, comma, numbers,sort \& compress, super}
\usepackage{appendix}
\usepackage{dcolumn}
\usepackage{bm}
\usepackage[colorlinks=true, linktocpage=true, linkcolor=blue, citecolor=blue,]{hyperref}
\usepackage[a4paper]{geometry}
\begin{document}
\title{Simultaneous Estimation of Elliptic Flow Coefficient and Impact Parameter in Heavy-Ion Collisions using CNN}
\author{Praveen Murali}

\author{Sadhana Dash}
\email{Sadhana.Dash@cern.ch}
	\affiliation{Department of Physics, Indian Institute of Technology Bombay, Mumbai, India-400076}
    \author{Basanta Kumar Nandi}
	\email{basanta.nandi@cern.ch}
	\affiliation{Department of Physics, Indian Institute of Technology Bombay, Mumbai, India-400076}

\begin{abstract}
A deep learning based method with Convolutional Neural Network (CNN) algorithm is developed for simultaneous determination of the Elliptic Flow coefficient ($v_{2}$) and the Impact Parameter in Heavy-Ion Collisions at relativistic energies. The proposed CNN is trained on Pb$-$Pb collisions at $\sqrt{s_{NN}}$ = 5.02 TeV with minimum biased events simulated with the AMPT event generator. A total of twelve models were built on different input and output combinations and their performances were evaluated. The predictions of the CNN models were compared to the estimations of the simulated and experimental data. The deep learning model seems to preserve the centrality and $p_{T}$ dependence of $v_{2}$ at the LHC energy together with predicting successfully the impact parameter with low margins of error. This is the first time a CNN is built to predict both $v_{2}$ and the impact parameter simultaneously in heavy-ion system.
\end{abstract}
\maketitle
\section{Introduction} 

The ultra-relativistic nucleus-nucleus collisions act as a unique tool to produce and probe the hot and dense matter created in laboratories at RHIC and LHC energies under controlled conditions \cite{alice, heavyion1, heavyion2}. The size of the dense and deconfined matter produced in such collisions depends on the impact parameter of the collision. This can be viewed from a simple picture of geometrical overlap of ions \cite{glauber}. However, the impact parameter which characterizes the initial state of the collisions is not directly measurable in experiments. It is generally estimated using experimentally measurable observables like charged particle multiplicity, transverse energy etc. which are strongly correlated with the impact parameter \cite{centrality}.
Flow is an experimentally relevant observable that provides information on the initial conditions of the collision, the equation of state, and the transport properties of the matter created in heavy-ion collisions \cite{flow, flow1, flow2}. The initial asymmetry in the geometrical overlap of the colliding nuclei as well as the presence of interaction between the constituents of the created matter causes the azimuthal anisotropy in the particle production. This anisotropy is one of the cleanest signatures of the collective flow in heavy-ion collisions. The widely measured flow coefficient is the elliptic flow coefficient ($v_{2}$), defined as the second Fourier coefficient of the azimuthal anisotropy of the produced particles. 

Recently, deep learning methods have been used in various scientific research challenges due to their ability to perceive unique features and patterns and to solve unconventional problems such as classification, regression, and clustering \cite{dnn}. There are even attempts to implement various machine learning algorithms to estimate $v_{2}$, impact parameter ($b$) etc. as well \cite{dnnflow, impact1, impact2}.
The present work aims to estimate the impact parameter and elliptic flow coefficient simultaneously in Pb$-$Pb collisions at $\sqrt{s_{NN}}$ = 5.02 TeV using the multi-transport AMPT model \cite{ampt}. A CNN-based regression model has been used to study the centrality as well as the $p_{T}$ dependence of the elliptic flow coefficient. The impact parameter has also been successfully estimated simultaneously using the same model.

\section{The AMPT Model}
A multi-phase transport model (AMPT) is a hybrid Monte Carlo model to study  p$-$A and A$-$A collisions at relativistic energies. It consists of four major components namely, the initial conditions, partonic interactions, hadronization, and subsequent interactions among the produced hadrons. The spatial and momentum distributions of the mini-jet partons and soft string excitations defining the initial state is obtained from the HIJING model \cite{hijing}. The partonic scatterings which include two-body scatterings only is described by the Zhang's parton cascade model \cite{zpc}. 
The default mode of AMPT uses the Lund string fragmentation model for hadronization while the string melting mode uses the quark coalescence model of recombining the quarks to form hadrons. The hadronization is followed by the interactions among the hadrons defined by a hadron cascade using a relativistic transport model(ART) \cite{art}. The hadronic interactions are stopped after a certain cut-off time and events are fetched subsequently.

\section{Target Observables}
In this section, the two observables, namely the elliptic flow coefficient, $v_{2}$ and the impact parameter, $b$ are discussed. 

\subsection{Elliptic Flow ($v_{2}$)}
In heavy-ion collisions, elliptic flow is an experimental observable which quantifies the azimuthal asymmetry of the momentum distribution of produced particles in the transverse plane. The phenomenon is 
observed due to the anisotropic pressure gradient in initial stages as a result of spatial asymmetry of the nuclear overlap region in non-central collisions. It can be expressed conveniently as the Fourier decomposition of the differential momentum distribution of particles in an event as the following, 

\begin{equation}
    \frac{dN}{d\phi} = \frac{1}{2\pi} \Biggr[1 + {\sum^{\infty}_{n=1}} 2 v_n \cos{(n(\phi - \psi_n))} \Biggr]
\end{equation}

Here, $v_{n}$ is the $n^{th}$ order harmonic flow coefficient, $\phi$ is the azimuthal angle and $\psi_n$ is the $n^{th}$ harmonic symmetry plane angle. As more interactions lead the system to thermalize, the $v_{2}$ values also probe the degree of thermalization in the system.  By construction, $v_{2}$ requires the information of the reaction plane angle on an event-by-event basis, whose measurement is non-trivial in experiments. There are a couple of methods that offer the solution such as the the cumulant method \cite{cumulant} and the usage of principal component analysis \cite{pcaflow}.
\subsection{Impact Parameter ($b$)}
Heavy-ions, unlike protons are extended objects and therefore, the properties of the created system as a function of the degree of overlap 
of the colliding nuclei are quite different. To study the properties of the created system, the collisions are classified into different centrality classes defined by the impact parameter of the collision. However, the impact parameter cannot be directly measured experimentally. The collision centrality is therefore estimated from the measured particle multiplicities, with the fair assumption of the multiplicity being linearly related to $b$. Phenomenologically, the total particle production was found to scale with the number of participating nucleons and the number of binary collisions. A realistic description of the nuclear geometry in a Glauber calculation could be used to relate these number of participating nucleons and binary collisions to the impact parameter, $b$ \cite{glauber}.

\section{Deep Learning with Convolutional Neural Network (CNN)} 

\subsection{Concept}

The computing machines require a specific algorithm to perform a task in which the solution to the problem is written in a top-to-down approach and the control flows accordingly resulting in an outcome. Yet, most of the problems come with no standard predefined set of rules to develop the algorithm that can solve them. Another direction is the high-complexity non-linear problems, where linear-based numerical methods usually fail. In such cases, Machine Learning (ML) with smart algorithms such as the Boosted Decision Trees (BDT)\cite{bdt}, Deep Neural Network (DNN) \cite{dnn}, Generative Adversarial Networks (GAN) \cite{gan} etc. could help the machine learn from the data through a process called training. ML is the branch of Artificial Intelligence (AI) that gives the ability to the computers to learn correlations from data components. Recently, DNN models have been used to map complex non-linear functions by using simulated data in the field of astronomy \cite{mlastro}. This ability could be exploited to train machine learning models to look for the hidden physics laws that govern particle production, anisotropic flow, spectra, impact parameter etc. in heavy-ion collisions. 
CNN, in general, have the following components,

\begin{enumerate}
    \item Convolution operation - A kernel-matrix is slid across the input map to produce a lower (or same) dimensional representation of the features. It is these kernels that get updated during training. Since the same kernel convolves all the pixels of the image, the CNN picks up spatial correlations inherent in the input image.

    \item Max/Average Pooling - The main purpose of using a Pooling layer is to reduce the dimension of the input image. The input image is divided into $2 \times 2$ windows, where the largest (or average) value is chosen and taken to the next step. In general, a $N \times N$ input image becomes $\frac{N}{2} \times \frac{N}{2}$ after one pooling step

    \item Flattening - All the pixels of the input features are flattened to form one long vector as it is easier to manipulate to get the desired output shape.
\end{enumerate} 

\section{Analysis Method}

The AMPT event generator was used to generate 50K minimum bias Pb$-$Pb events at $\sqrt{s_{NN}} = 5.02$ TeV. The string melting version of AMPT ISOFT mode was used. All charged particles with transverse momentum ($p_{T}$), $0.2 < p_T < 5.0$ GeV/{\it c} and a pseudorapidity acceptance of $|\eta| < 0.8$ were considered for the estimation of elliptic flow coefficient using the event plane method. In AMPT simulation, the reaction plane angle ($\psi_{n}$) was set to zero, although it is non-trivial in experiments. One can therefore, obtain the elliptic flow coefficients as $v_2 = \langle \cos{(2\phi)} \rangle$. The average was taken over all the selected charged particles in an event.  80\% (40K) of the total event sample was used for training and validation while 20\%  was used for testing and evaluation. The analysis was performed using Keras v2.11.0 Deep Learning API \cite{keras} with Tensorflow v2.11.0 \cite{tensorflow}.

\subsection{Input and Output}

\begin{figure*}    
 \centering   \includegraphics[height=45mm]{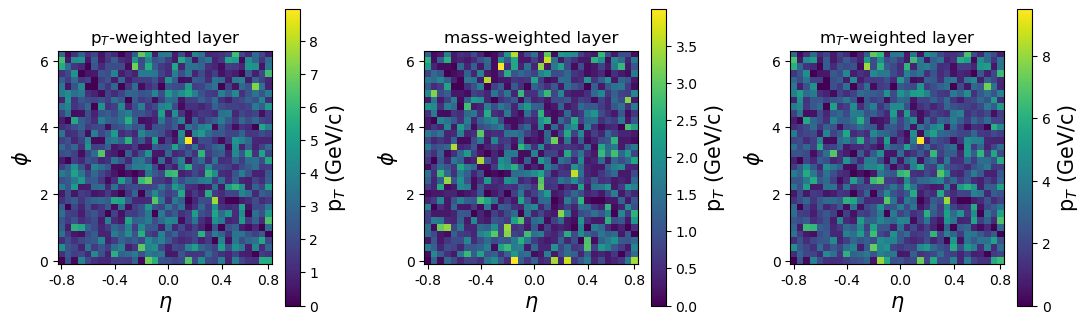}
    \centering
    \caption{The three layers of the weighted $(\eta, \phi)$ space with (32, 32) bins for a single Pb$-$Pb collision at $\sqrt{s_{NN}} = 5.02$ TeV.}
    \label{fig1}
\end{figure*}

In this study, a CNN-based regression model was trained for the target observables $v_{2}$ and $b$. The binned $(\eta-\phi)$ coordinate space for all charged particles in an event was considered for the primary input space. Here, $|\eta| < 0.8$ and $\phi \in [0, 2\pi]$. The bin settings, (16, 16) and (32, 32) was considered for this procedure. This selection was chosen because anything below 16 offered very little number of features for the CNN to work with and anything above 32 allowed large sparse inputs where most of the features were absent. It is always a good practice to work with powers of two to make it easy for the machine to compute. Additional kinematic information was included as two secondary layers to the $(\eta-\phi)$ space. These two layers were simply weighted by the $p_{T}$ and the mass of the charged particles. From these two input layers, a third layer, containing the information of $p_T$ and mass, was constructed. This layer was weighted by the transverse mass ($m_{T}$) of the particles defined by $m_{T} = \sqrt{p_{T}^2 + m^2}$. Figure \ref{fig1} shows the three layers of the weighted $(\eta-\phi)$ space with (32, 32) bins for a single event. The $(\eta-\phi)$ space was directly read as images in the CNN.   

A total of four different options for input, (only $p_{T}$, only $m$, both ( $p_{T}$ and $m$), and $m_{T}$), and three different options for output, (only $v_{2}$, only $b$, and both ($v_{2}$ and $b$)) were considered.  All these different combinations were addressed by considering twelve (4 inputs x 3 outputs) separate models. A single CNN architecture with variable input and output neurons was used. 
The impact parameter values were  normalized such that $b \in [0, 1]$
in order to make the estimations more machine-friendly. Such a normalization step was not required for $v_{2}$ since $v_{2} \in [0, 1]$. None of the inputs were normalized considering that this model could be directly implemented to data measured at LHC experiments.

\subsection{CNN Architecture}


For the present study, the input was either $p_{T}$-weighted, mass-weighted or $m_{T}$-weighted or both $p_{T}$ and mass-weighted $\eta - \phi$ distribution. Hence, the shape of the input layer was (16, 16, 1) for (16, 16) bins weighted by $p_{T}$ or mass or $m_{T}$, (16, 16, 2) for (16, 16) bins weighted by both $p_{T}$ and mass, (32, 32, 1) for (32, 32) bins weighted by $p_{T}$ or mass or $m_{T}$ and (32, 32, 2) for (32, 32) bins weighted by both $p_{T}$ and mass.  All the twelve models were constructed with the same architecture to ensure uniformity while comparing the performance of these models.
The first and the second convolutional(Conv) layer had 16 filters. The third and the fourth convolutional layer had 32 filters. Furthermore, the output of the fourth convolutional layer was flattened to a multi-layer perceptron (MLP). This MLP led to two dense layers with 64 and 32 neurons, respectively. Finally, the neural network ended with the output neuron(s). The output was either $v_{2}$ only or $b$ only or both ($v_{2}$ and $b$).
\\\\
The activation used for every Conv and Dense layer was linear. The padding of each Conv layer was kept to be same to avoid over-complicating the shape of the layers. To avoid over-fitting and generalization of the learned pattern mapping, the following three techniques were used. 

\begin{enumerate}
    \item Each Conv layer was followed by a Group Normalization technique with the number of groups set to four.

    \item Conv layer 2 and 4 were followed by a dropout layer with a dropout rate of 0.1.

    \item Every Conv and Dense layer had an L1L2 kernel regularizer.
\end{enumerate}

The model was compiled with a Mean-Squared Error loss function and the Adam algorithm was used to update the kernel weights. The Mean-Absolute Error (MAE) was also tracked while training to evaluate the learning. The model trained with a batch size of 256 for 70 epochs showed a good convergence with no over-fitting.

\section{Results and Discussion}
\subsubsection{General Performance}
Table \ref{table1} shows the performances of all the 24 models that were built and trained for the 2 bin settings, (16 $\times$ 16) and (32 $\times$ 32). The MAE column shows the error obtained while testing the models. The percentage indicates the magnitude of the error but not the relative error. It can be observed that all 24 models performed well with maximum error being less than 6\%. However, a detailed discussion predominantly focusing on the bin setting (32, 32) is given in the following section as bin (32, 32) outperformed (16, 16) in several cases.

\begin{table}[h!]
\centering
\begin{tabular}{|c|c|c|c|}
\hline
\textbf{Input} & \textbf{Output} & \textbf{MAE} & \textbf{MAE} \\ 
&      &  (16 $\times$ 16) &  (32 $\times$ 32) \\ \hline 
p$_T$     & v$_2$    & 5.30\% & 4.93\% \\ \hline
p$_T$     & b     & 4.12\% & 3.83\% \\ \hline
p$_T$     & both  & 4.73\% & 4.43\% \\ \hline
mass   & v$_2$    & 5.63\% & 5.46\% \\ \hline
mass   & b     & 4.47\% & 3.39\% \\ \hline
mass   & both  & 4.99\% & 4.48\% \\ \hline
$m_{T}$ & v$_2$    & 5.38\% & 5.28\% \\ \hline
$m_{T}$ & b     & 3.84\% & 3.90\% \\ \hline
$m_{T}$ & both  & 4.50\% & 4.48\% \\ \hline
both   & v$_2$    & 5.52\% & 4.77\% \\ \hline
both   & b     & 3.86\% & 4.43\% \\ \hline
both   & both  & 4.67\% & 4.37\% \\ \hline
\end{tabular}
\caption{Model performance comparison with different inputs and outputs.}
\label{table1}
\end{table}
\subsubsection{Mean Absolute Error}

\begin{figure*}
\centering
\includegraphics[height=90mm]{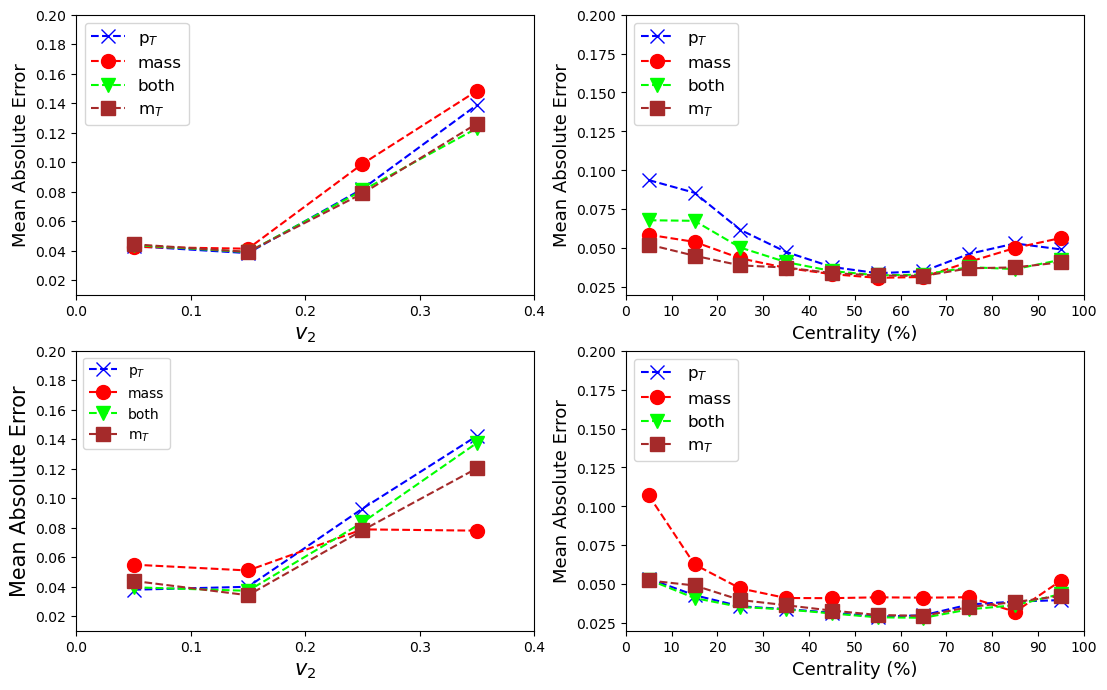}
    \centering
    \caption{The variation of MAE as a function of the target observables for all 24 models.}
    \label{fig2}
\end{figure*}

Figure \ref{fig2} shows the variation of MAE for different values of target observables and therefore compares the performance of all the 12 models under the bin setting of (32, 32). The top left panel describes the model where the output is $v_{2}$ while the top right one describes the model where the output is $b$. The bottom two panels describe the model with both outputs, $v_{2}$ and $b$ plotted separately. In each figure, the blue curve describes the model where the input was a $p_{T}$-weighted image, the red curve describes the model where the input was a mass-weighted image, the orange curve describes the model where the input was $m_{T}$ weighted image and the green curve describes the model where the inputs were both $p_{T}$ and mass-weighted images. 
\\\\
As evident from the figure, in predicting $v_{2}$, the CNN models performed well for $v_{2} \leq 0.4$. The MAE only increased beyond the mentioned value and can be attributed to fewer statistics in this region. But the interesting point is that despite low statistics, the CNN models could learn the pattern within a relative error ($\Delta v_{2} / v_{2}^{true}$), of 5\%. Models with different inputs performed very similarly.
\\\\
The prediction of $b$ is slightly different. The models seemed to perform well for mid-central regions ( (30 - 70 )\%) in comparison to central and peripheral collisions. Again, this behavior could be attributed to less number of samples in these regions, thereby affecting the learning.  Moreover, the worst performing class, (0-10)\% centrality, only deviated by 0.25 fm with a relative error, $(\Delta b / b^{true})$, of 7.1\%. In the best scenario, the relative error is as low as 0.5\% which is impressive.
\\\\
When it comes to inputs, the CNN models have a preference. In the case where the output was only $b$, the model with only $p_{T}$-weighted image input performed poorly while the model with only $m_{T}$-weighted input performed best. In the case where the output was both $v_{2}$ and $b$, the model with only mass-weighted image input performed poorly while the rest of the models performed more or less equally well. However, the differences in the performance was due to the contribution from low statistics regions and yet they were not significant in terms of relative error.

\begin{figure*} 
\centering
\includegraphics[height=100mm]{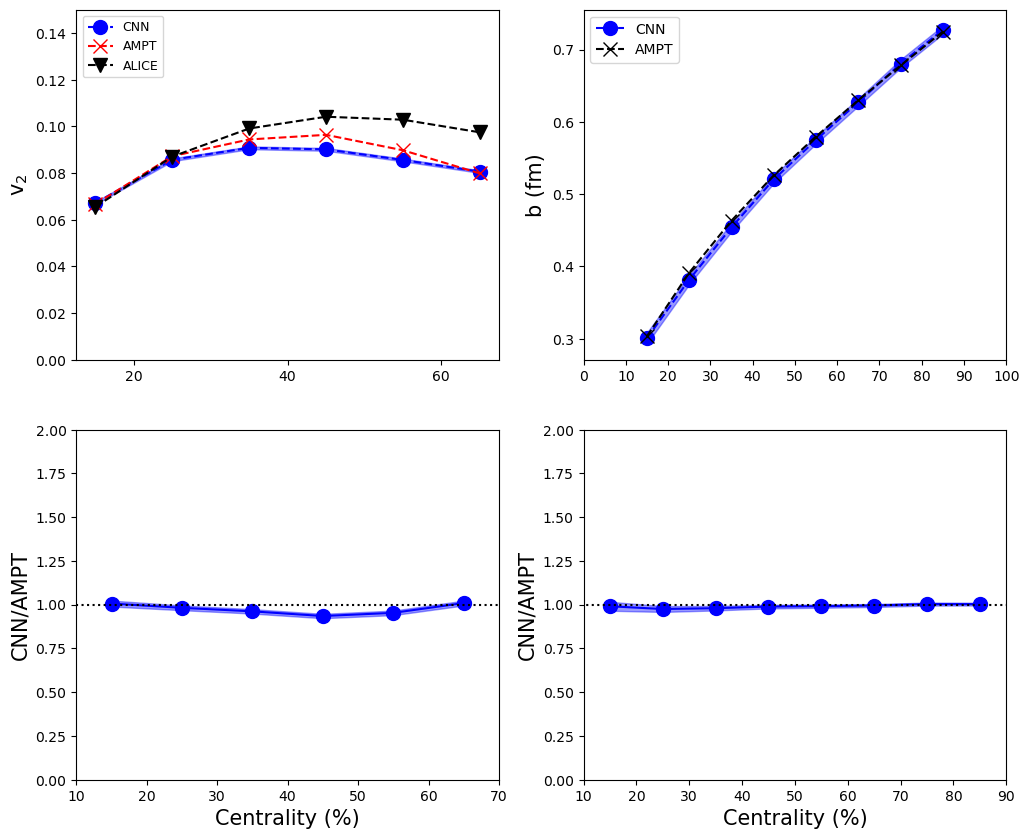}
    \centering
    \caption{Centrality dependence of the target observables for the model where the input has both $p_{T}$ and mass-weighted layers. The ALICE data is taken for Pb$-$Pb collisions at $\sqrt{s_{NN}}$ = 5.02 TeV \cite{aliceflow}.}
    \label{fig3}
\end{figure*}

Figure \ref{fig3} shows the centrality dependence of the target observables predicted with the model with $p_{T}$ and mass-weighted $(\eta - \phi)$ distribution as inputs. The top figure of the left panel shows the centrality dependence of $v_{2}$ and is compared with the experimental data \cite{aliceflow} while the figure on the right shows the centrality dependence of $b$. The bottom panels show the ratio of CNN and AMPT predictions quantifying the agreement between the model and the CNN prediction. 
\\\\
As can be seen, the predictions of the CNN model has good agreement with the simulation data for all centrality classes. The ratio is close to one in both the cases. When compared to different (input, output) combinations, this model seems to perform well. This can be attributed to input having two layers giving maximum kinematic information. By training the CNN model with minimum bias Pb$-$Pb collisions at $\sqrt{s_{NN}} = 5.02$ TeV, one allows the machine to learn physics for a larger and more complex system. This model can be directly implemented in heavy-ion collision systems tuned to LHC energies. 
\\\\
However, there are some discrepancies in the peripheral regions. It is to be noted that the estimation of $v_{2}$ in ALICE using $|\Delta \eta| > 1$ has some level of non-flow contributions. In addition, different methods of flow estimation could also introduce a degree of uncertainty. This is the first time a single model is employed to estimate both $v_{2}$ and $b$. 

\subsubsection{$p_{T}$ dependence of $v_{2}$}
\begin{figure*}
\centering
    \includegraphics[height=70mm]{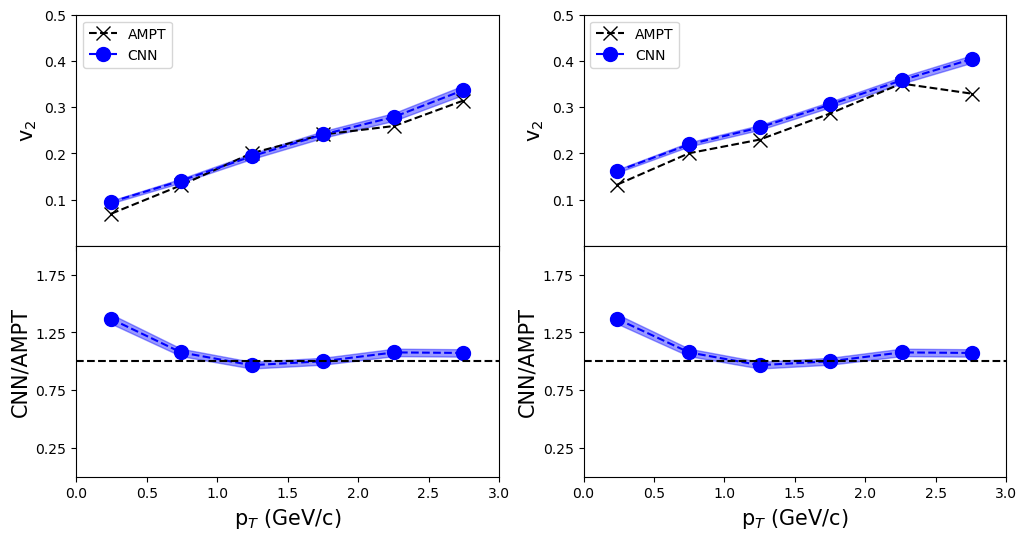}
    \centering
    \caption{ Variation of $v_{2}$ with $p_{T}$. The left panel shows the model with input $m_{T}$-weighted in (16,16) bins while the right panel shows the same with (32,32) bins.}
    \label{fig4}
\end{figure*}

 Figure \ref{fig4} shows the variation of $v_{2}$ with $p_{T}$ for different centrality classes. The best model for each, (16,16) and (32,32), bin is shown. Usually, fewer particles are detected for high $p_{T}$, suggesting that the input regions for high $p_{T}$ would be very sparse. Despite this constraint, all twelve models have predicted the relation in both bin settings. This feature along with the centrality dependence make these models even more robust and versatile.

\subsubsection{Activation Maps}
In the case of convolutional neural networks (CNNs), commonly used in visual recognition tasks, various methods have been developed to visualize their behavior. One such method is class activation mapping (CAM) \cite{cam}, which involves global average pooling and calculating the importance of neurons within a CNN. This technique allows for the localization of crucial regions in two dimensional matrices that contribute to the model's success in classification tasks. Building on the CAM method, Selvaraju et al. introduced a more generalized approach called gradient-weighted class activation mapping (Grad-CAM) \cite{gradcam}. Grad-CAM can be applied to a wider range of CNN architectures and provides deeper insights into the learned features of the neural network.
\\\\
In the CAM method, the class activation map for
class c (in the present case, c is trivial and can be suppressed because we have a regression problem instead of a classification problem) is defined as $M^c$.

\begin{equation}
    M^c_{x,y}(A) = {\sum_k} \omega^c_k f_k(A;x,y) 
\end{equation}

Here, $f_k(A;x,y)$ represents the activation of the feature k in the final convolutional layer at spatial location $(x,y)$ of input data $A$, and $\omega^c_k$ is the weight measuring the importance of unit k for class c. In Grad-CAM, $\omega^c_k$ is defined as the result of performing global average pooling on the gradient of the score for class c for activations $f_k(A;x,y)$:

\begin{equation}
    \omega^c_k = \frac{1}{Z} \mathbf{\sum_x} \mathbf{\sum_y} \frac{\partial y^c}{\partial f_k(A;x,y)}
\end{equation}

where $1/Z \mathbf{\sum_x} \mathbf{\sum_y}$ is the operation of global average pooling. Subsequently, the gradient-weighted class activation map is given as

\begin{equation}
    L^c = \mathbf{\sum_k} \omega^c_k f_k(A;x,y)
\end{equation}

\begin{figure*}
\centering
    \includegraphics[height=80mm]{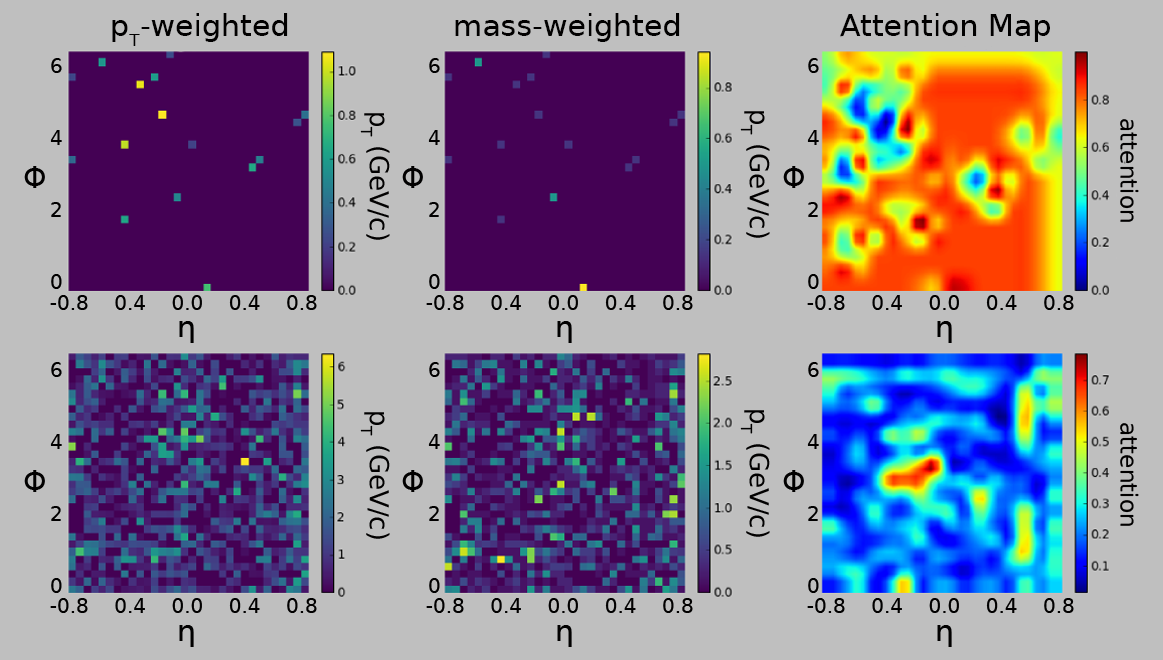}
    \centering
    \caption{Attention Maps for two different input space distributions.}
    \label{fig5}
\end{figure*}

The {\bf Attention Maps} for the best-performing model was obtained. 
Figure \ref{fig5} shows the  Attention Maps for two samples with different input distributions. It can be observed that the region of the input space where very little information is available, demands maximum attention from the CNN. The top figure with sparsely distributed information has a bright red (high attention) heat map while the bottom plot with uniformly distributed information has a blue (low attention) heat map. 

\begin{figure*}
\centering
    \includegraphics[height=100mm]{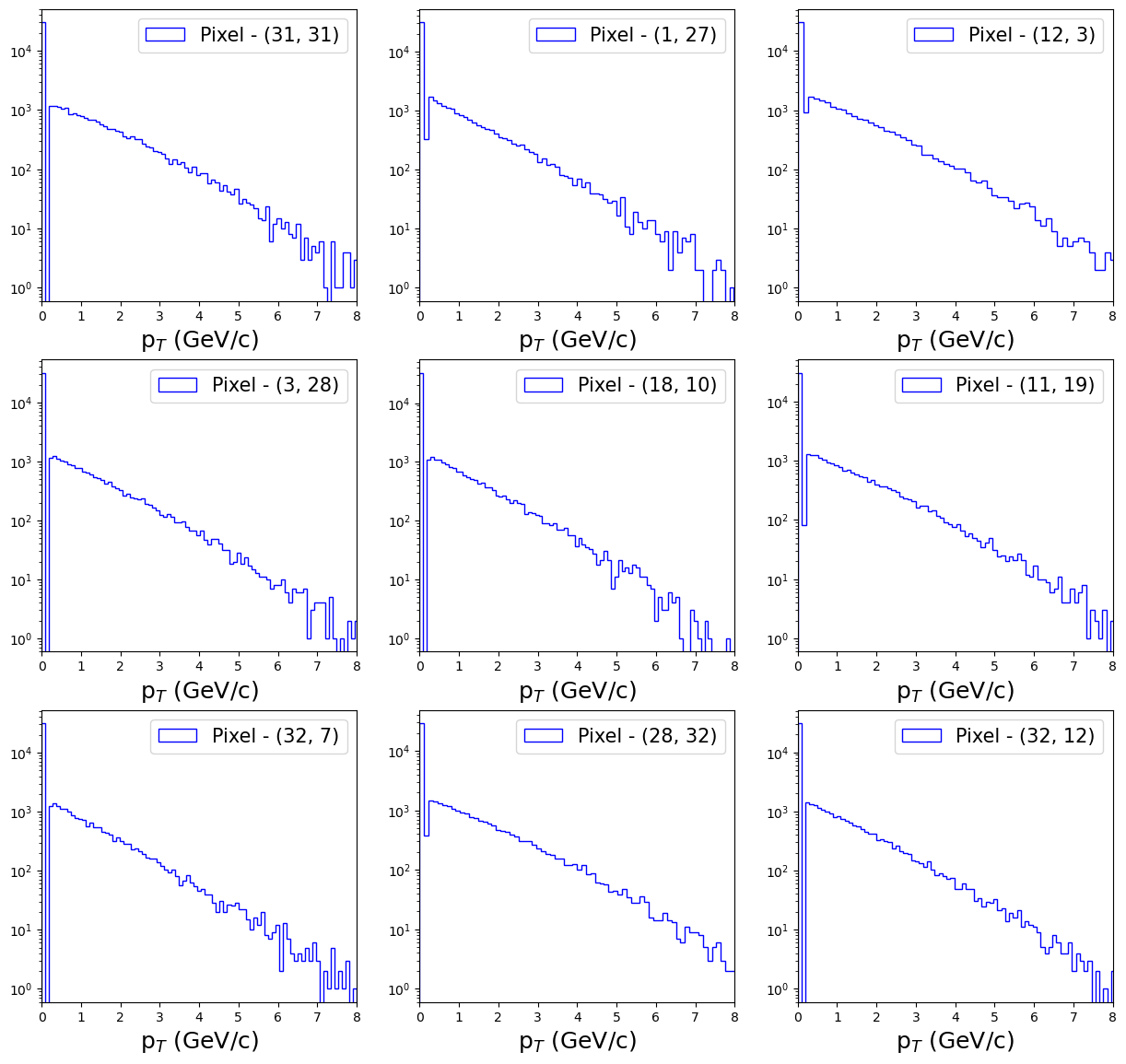}
    \centering
    \caption{Distribution of pixel values for $p_{T}$-weighted layers of 9 random pixels across the events.}
    \label{fig6}
\end{figure*}

This may be surprising since CNN, or any neural networks for that matter, naturally tend to map regions of information to target variables. However, the attention maps point towards a different story.  Figure \ref{fig6} shows the distribution of pixel values for $p_{T}$-weighted layers across all the events. It is seen that the overwhelming majority of the pixel values lie very close to zero. This directly shows that most of the data are sparsely distributed inputs. The y-axis is log-scaled to show the sheer difference in magnitude of the number of pixels with values close to zero from those that are non-zero. The shape of the distribution is the same for all pixels. Similar kind of behavior was also seen for mass-weighted layers. This indicates that the CNN is predominantly trained on samples with highly localized regions of information. Inputs that are uniformly distributed are essentially rare and the CNN sees fewer of them while training. So, it does make sense for the CNN to effectively rely on regions of less information to extract target variables.

\subsubsection{Quality over Quantity}
Neural nets usually rely on a large dataset to effectively extract mappings between input and output neurons. A sophisticated network like the one built in this work, performs well in regions of fewer statistics like for $v_{2} > 0.2$ and highly central ( or peripheral) collisions. However, one can always train these models with newer datasets to further improve their accuracy. Since these models are already performing efficiently, newer datasets must exhibit good quality rather than a large quantity. Quality is subjective but it can be defined precisely for a given ML problem. In the case of weighted $\eta - \phi$ distribution as input, the quality of a dataset can be defined as the degree to which the distribution of the new dataset agrees with the one that was used for training. Any dataset that has a pixel distribution that is similar to Figure \ref{fig6} is good. The CNN, when trained on such datasets, will certainly learn new patterns and will improve its performance. Deviations from the distributions of pixel values shown in Figure \ref{fig6}  will result in a dip in the performance of the CNN. 
\\\\
However, datasets from other event generators which incorporated inbuilt flow like HYDJET++ \cite{hydjet} can also be used for training but for those datasets, the quality matters over quantity. In fact, to test the robustness of the models built in this work, one can generate data from different event generators and analyze the performance of the predictions. The $(\eta-\phi)$ distributions from the experimental data can be directly used. Since these distributions would have to most realistic distribution of pixel values, even a small dataset of 10K events should give a good understanding of how inputs must be distributed for the CNN to learn best. Such comparison would also provide some room for improvement in these Monte Carlo based event generators.

\section{Conclusion}
In conclusion, the implementation of a novel CNN technique to include the kinematic properties of charged particles produced in heavy-ion collisions is carried out to estimate the elliptic flow coefficient and the impact parameter simultaneously. The proposed CNN model uses the two-dimensional $\eta - \phi$ distribution of charged particles weighted with $p_{T}$, mass and $m_{T}$ as model input under two different bin settings, (16, 16), (32, 32). A total of 12 different combinations were built, trained and tested with minimum-bias Pb$-$Pb collision events simulated at $\sqrt{s_{NN}} = 5.02$ TeV utilizing the AMPT event generator. The CNN models show a good agreement between the predictions and the simulated values. The best performing combination was obtained for both $p_{T}$ and mass-weighted $(\eta-\phi)$ input distribution with outputs as $v_{2}$ and $b$ under the bin setting of (32, 32). This particular combination utilized the maximum available information to predict the quantities with the best precision. Furthermore, all the models seem to preserve the $p_{T}$-dependence of $v_{2}$ across all centrality classes.

\section{Acknowledgement}
    S.D would like to acknowledge the SERB Power Fellowship, SPF/2022/000014 for the support in this work.

\end{document}